# Random Lasers for Broadband Directional Emission

Sebastian Schönhuber[1*], Martin Brandstetter[1*], Thomas Hisch[2], Christoph Deutsch[1], Michael Krall[1], Hermann Detz[3], Aaron M. Andrews[4], Gottfried Strasser[4], Stefan Rotter[2], Karl Unterrainer[1]

[1]Photonics Institute and Center for Micro- and Nanostructures, TU Wien, 1040 Vienna, Austria
[2]Institute for Theoretical Physics, TU Wien, 1040 Vienna, Austria
[3]Austrian Academy of Sciences, Dr. Ignaz Seipel-Platz 2, 1010 Vienna, Austria
[4]Institute of Solid State Electronics and Center for Micro- and Nanostructures, TU Wien, 1040 Vienna, Austria
*Corresponding author: sebastian.schoenhuber@tuwien.ac.at, and martin.brandstetter@tuwien.ac.at

**Broadband coherent light sources are becoming increasingly important for sensing and spectroscopic applications, especially in the mid-infrared and terahertz (THz) spectral regions, where the unique absorption characteristics of a whole host of molecules are located. The desire to miniaturize such light emitters has recently lead to spectacular advances with compact on-chip lasers that cover both of these spectral regions. The long wavelength and the small size of the sources result in a strongly diverging laser beam that is difficult to focus on the target that one aims to perform spectroscopy with. Here, we introduce an unconventional solution to this vexing problem relying on a random laser to produce coherent broadband THz radiation as well as an almost diffraction limited far-field emission profile. Our random lasers do not require any fine-tuning and thus constitute a promising example of practical device applications for random lasing.**

Various spectroscopic techniques rely on the specific absorption features of numerous molecules within the mid-infrared and terahertz (THz) spectral regions, which allow their unambiguous identification. Although broadband coherent sources of radiation are already available within these frequency regions [1–5], a compact size and electrically pumped operation are additionally desired for actual applications. These criteria are fulfilled by quantum cascade lasers (QCLs), semiconductor sources which are able to provide broadband emission at mid-infrared [6,7] and THz [8,9] frequencies. However, for typically used THz QCL waveguides the output aperture is on the order of 10 µm, while the emission wavelength is about 100 µm. This, in fact, leads to a very divergent output beam, which is additionally distorted by interference effects [10,11]. Since external optical elements such as lenses or antennas are difficult to handle on a large scale [12,13], several monolithic concepts have been pursued to address this issue. To improve the far-field for facet emission, e.g., a 3rd order distributed feedback (DFB) grating can be used [14]. Another approach is to coherently emit from a large area on the laser surface. In contrast to interband semiconductor lasers, QCLs can only generate in-plane radiation due to intersubband selection rules, preventing the realization of VCSEL-type resonators. Thus several concepts have been developed to couple out the in-plane cavity mode in vertical direction, including 2nd order DFB gratings [15,16], and photonic crystal (PhC) cavities [17]. Recently, also non-periodic resonator structures such as graded photonic heterostructures [18], or quasicrystal cavities [19,20] have been developed. However, all on-chip techniques providing surface emission demonstrated so far are designed to support a single laser mode only, while for many spectroscopic applications a broadband coherent light source is necessary. To achieve the objective of a broadband and at the same time very collimated laser light emission in the THz regime, we propose here a radical break with previous attempts to carefully engineer the laser surface and to rely, instead on random laser cavities [21] supporting multiple laser modes within the laser gain bandwidth. In spite of the random nature of light scattering in these structures, we demonstrate almost diffraction limited broadband surface emission with this concept.

For the experimental realization of this THz random laser, a 10 µm thick epitaxially grown quantum cascade active region was used as the gain medium. Due to their unipolar operation principle, QCLs do not suffer from surface recombination enabling almost arbitrary cavity geometries without any influence on the electrical properties. A double-metal waveguide [22] was used for vertical mode confinement, where the generated radiation is guided between two metal layers within the active region. Randomly placed, non-overlapping scattering elements were realized by etching vertical holes through the complete semiconductor structure. The diameter was chosen to be 20 µm, which is on the order of the emission wavelength in the medium, schematically shown in Fig. 1(a). Different scattering configurations have been realized with 8% to 34% filling fraction (determined by the area of the holes with respect to the area of the entire circular cavity). The holes define the in-plane random laser modes and additionally couple out the radiation in surface direction. A scanning electron microscope (SEM) picture of a typical device is shown in Fig. 1(b). For periodic and quasi-periodic cavities [17,20] typically an absorbing boundary is implemented to exclude effects of the device edges on the desired cavity mode. This, however, is not necessary here, since the cavity is supposed to support multiple random modes rather than a particular one. The lasers are electrically contacted using a wire bonding technique (not shown here), which only has a minor influence on the laser's far-field emission [17]. Measurements show that the device performance, which is typically represented by the maximum operation temperature for THz QCLs, is decreasing with increasing scattering density, see Fig. 1(d). Since each hole contributes to the extraction of the radiation from the laser cavity, a larger filling fraction (i.e. more holes) leads to increased output losses. This effect can also be observed in the maximum output power, which is increasing for higher filling fractions, leading to peak output powers of more than 40 mW measured in vertical direction, see Fig. 1(c),(d). This behavior is different from previously studied random lasers, where an increased scattering density primarily leads to a higher Q-factor [23,24].

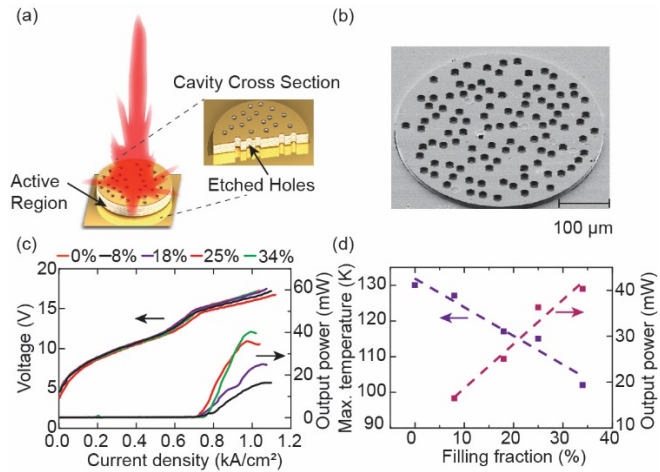

**Fig. 1.** Experimental realization. (a) Schematic of the random laser providing directional surface emission. (b) SEM picture of a typical device with 25% hole-filling fraction and a total resonator diameter of 500 µm. The diameter of the holes is 20 µm. (c) Current-voltage (IV) and light-current (LI) characteristics of devices with various filling fractions, measured at a heat-sink temperature of 5 K. Peak output powers of more than 40 mW emitted in surface direction have been achieved. The reference IV-curve of a device without scattering elements (0%), supporting whispering gallery modes, shows the same behavior and thus confirms that the electrical characteristics are not influenced by the cavity structure. (d) Maximum operating temperature, representing the individual device performance, and peak output power measured in surface direction, indicating increased outcoupling losses for higher hole-filling fractions.

Measured far-fields and the spectral characteristics of devices with different filling fractions are depicted in Fig. 2. Devices with 18% filling fraction provide a beam with about Θ=7° full width at half maximum (FWHM) divergence, and additionally cover a frequency range of almost 400 GHz, corresponding to the total spectral gain bandwidth of the active region. Thus, the experiment indeed demonstrates that almost diffraction limited beams can be obtained, implying in-phase coherent emission from the laser surface leading to a Gaussian-like far-field due to constructive interference, similar to a 2nd order DFB laser. The distinctive feature of the presented random lasers compared to previously demonstrated concepts [14–18,20] is the fact that the far-fields consist of multiple spectral components. Although different scattering configurations with the same filling fraction show different emission patterns (see Fig. S1), a general trend is observable for the far-field to become degraded for higher filling fractions in terms of more strongly pronounced interference effects. This observation can be attributed to fewer lasing modes contributing to the far-field due to the higher outcoupling losses. A similar degradation of the far-field characteristics can also be observed for a reduced number of emitting holes (8% filling fraction) in Fig. 2(b). Similar to previous random laser experiments [25], mode splitting has been observed for low filling fractions (see e.g. black arrows in Fig. 2(a), 18% filling fraction), which is attributed to the coupling between individual spatially overlapping modes, occurring only for low filling fractions, where modes are very delocalized.

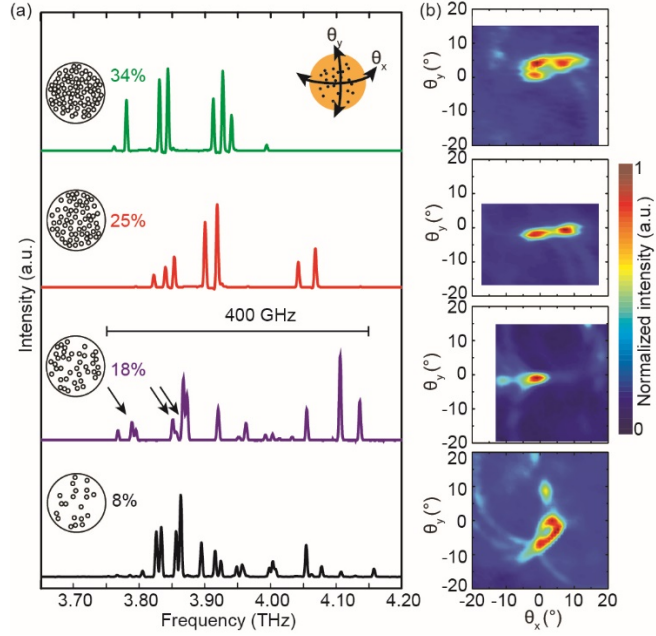

**Fig. 2.** Measured spectra and far-fields. The lasers were operated at the bias condition (pump strength) that provides the maximum output power. (a) Measured spectra for devices with different filling fractions. More lasing modes can be observed for lower filling fractions due to the lower outcoupling losses. Broadband emission with a bandwidth of almost 400 GHz is achieved for a filling fraction of 18%. The random laser cavity is illustrated schematically in the insets for each filling fraction. The black arrows indicate a spectral splitting of spatially overlapping modes. (b) Highly directional emission is observed in the direction perpendicular to the laser surface, with the best result obtained for devices with 18% filling fraction, showing about $\Theta=7°$ FWHM beam divergence, which is almost diffraction limited considering a 500 µm diameter output aperture and an emission frequency of 3.9 THz (resulting in 6.6° FWHM beam divergence).

To analyze the mechanism responsible for the observed collimated far-field, we first employ a simplified model in which the holes of the random laser are represented by randomly arranged and temporally coherent in-phase dipole emitters with a given polarization (orientation), see Fig. 3(a). Since our lasers operate in the multimode emission regime, the total far-field consists of contributions from several modes. This is taken into account by an incoherent superposition of the far-field intensities corresponding to different individual dipole orientations (representing the individual laser modes) in a spatial dipole arrangement, which is the same for all orientations (representing the random cavity). A diffraction limited beam is created, if all dipole emitters are aligned in the same direction (100% net-polarization), see bottom panel in Fig. 3(b). This scenario is similar to a 2nd order DFB laser, where the resonator structure shows the same periodicity as the lasing wavelength, which not only leads to coherent resonant feedback, but also to constructive interference of the radiation emitted from each individual scattering element in vertical direction. In the totally un-polarized (though temporally coherent) case with 0% net-polarization interference effects dominate the far-field pattern, see top panel in Fig. 3(b). In between these two limiting values a partial net-polarization still leads to constructive interference for a fraction of the emission perpendicular to the surface ($\Theta=0°$) regardless of the direction of the net-polarization. Calculations show that for an incoherent superposition of multiple far-fields, representing multimode emission, the components perpendicular to the surface add up, and interference effects become less pronounced. This gives rise to a very collimated emission in the direction orthogonal to the surface consisting of multiple spectral components, see middle panel Fig. 3(b). In Fig. 3(d), the calculated directionality, defined as the fraction of the radiation emitted within an angle of $\Theta=10°$ in this vertical direction, is shown versus the filling fraction for different degrees of the dipole net-polarization. A larger amount of coherently emitting elements leads to a more directional far-field, as consistent with the measurements on devices with 8% and 18% filling fraction. However, for higher filling fractions fewer lasing modes are observed in the experiment (25% and 34% filling fraction), which leads to more strongly pronounced interferences in the far-field. This effect is fully reproduced by the model when considering a smaller number of incoherently superposed far-fields, see Fig. 3(c). In the experiment, the number of lasing modes depends on the pump strength. As depicted in Fig. S2, the far-field quality is indeed improved for an increasing number of lasing modes. Thus, we attribute the experimentally observed optimum of the far-field quality for devices with 18% filling fraction to be determined by a suitable combination of a large number of lasing modes and a high density of emitting holes. Moreover, our model also shows that an in-phase dipole net-polarization leads to a linear far-field polarization of the vertically emitted radiation, see Fig. 3(e). Additional measurements that we performed on devices with 34% filling fraction indeed show that each spectral component of the far-field is mainly linearly polarized, see Fig. 3(f) and Fig. S3. As illustrated in Fig. S4, even adding a random but static temporal phase to each dipole emitter preserves the directional emission (albeit with a weaker linear polarization), as long as the dipoles have a net-polarization in the near-field.

For a more accurate description of the random laser cavities, full three-dimensional and vectorial finite-element (FEM) simulations have been performed for the device geometry illustrated in Fig. 4(a),(b). These calculations show that this structure supports multiple modes within a bandwidth of 400 GHz centered at 3.9 THz. The z-component of the electric field inside the cavity $|E_z|$ of one particular mode is shown in Fig. 4(c), which is distributed across the whole cavity. Despite the circular shape of the resonator the calculations did not yield whispering gallery modes, since these are inhibited by the holes. In analogy to the dipole-model, we also evaluate the near-field net-polarization $P_{Net}$ for these calculated 3D modes in the plane 1 µm above the top of the cavity (see Supplement 1). In Fig. 4(d), $P_{Net}$ is plotted versus the 10°-far-field directionality for various cavity modes, again showing that a near-field net-polarization is strongly correlated with a directional far-field. The grey circle indicates the particular far-field, plotted in Fig. 4(e), corresponding to the eigenmode shown in Fig. 4(c). To determine the net-polarization of modes from the 3D simulations on a quantitative level, we expect that taking into account non-linear interactions between the laser modes [26] as well as effects, which are specific to the employed quantum cascade structures [27] will be necessary. An incoherent far-field superposition of 20 modes with the highest $P_{Net}$ is depicted in Fig. 4(f), clearly displaying collimated multimode vertical emission.

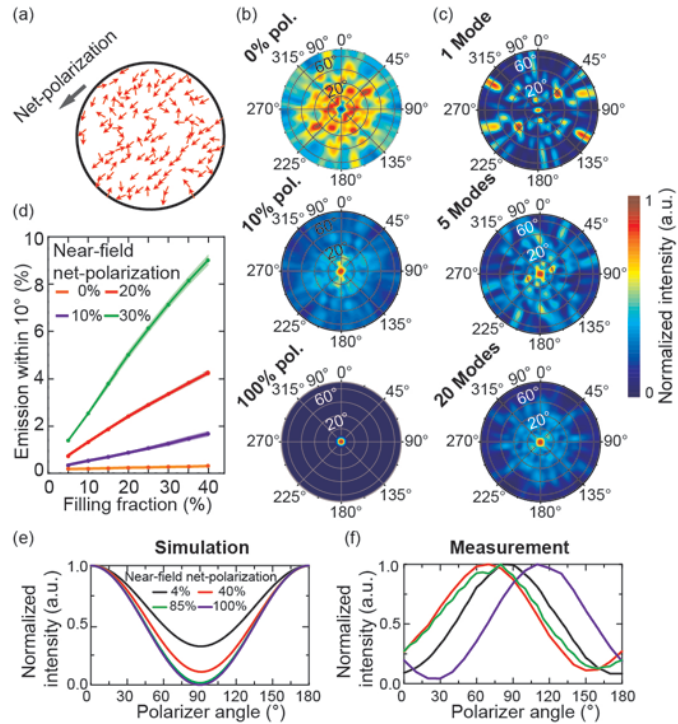

**Fig. 3.** Far-field calculations using a 2D geometry. (a) Our model consists of randomly arranged in-phase dipole-radiation sources with a defined orientation (polarization), indicated by the arrows. The large arrow indicates the near-field net-polarization of the ensemble of dipoles. (b) Far-field calculations of a device with 25% filling fraction for different net-dipole polarizations. 20 far-fields from different dipole orientations are incoherently superposed to model multimode emission. For a completely random dipole orientation (0%), interference effects dominate. Already a partial net-polarization (10%) leads to collimated emission in vertical direction. (c) Incoherently superposing multiple far-fields leads to less pronounced interference effects (25% filling fraction, 10% net-polarization). (d) The directionality increases for increasing filling fraction and increasing net-polarization (considering 100 scattering configurations with 20 superposed modes for each filling fraction and net-polarization). (e) Simulated far-field polarization within an angle of Θ=28° in the direction normal to the surface of a single mode for different near-field net-polarizations (filling

fraction 35%). (f) Measured far-field polarization of different laser modes (34% filling fraction, observation angle $\Theta=28°$ perpendicular to the surface), indicating linearly polarized vertical emission due to an in-phase near-field net-polarization.

In conclusion, we demonstrate a laser cavity incorporating randomly placed holes, acting as scattering elements, which determine multiple in-plane random laser modes and, moreover, couple out the radiation in surface direction. The frequency independent feedback mechanism and constructive interference in surface direction lead to a broadband collimated laser emission, as required for applications. Our concept may also be an interesting tool to investigate the complicated behavior of random lasers itself. By evaluating the far-field properties of the laser, conclusions about the cavity modes can be drawn, which are usually difficult to access in actual experiments [28]. Furthermore, the THz spectral region that we operate in provides the advantage of a large emission wavelength, leading to conveniently adjustable cavity structures, which allows us to exactly determine the geometry of the random laser and thus to engineer its emission characteristics in even more detail [29].

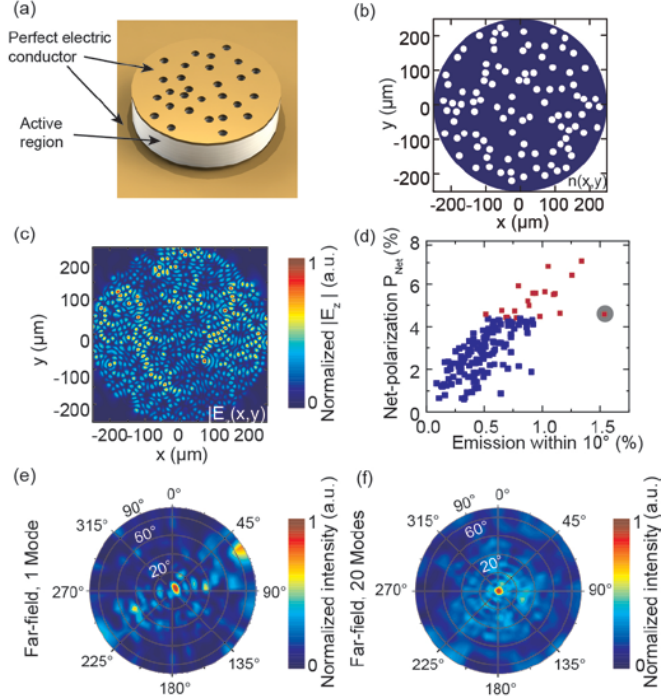

**Fig. 4.** 3D FEM simulations of the random laser cavity. (a) The simulation geometry consists of a dielectric layer representing the active region, which is located between two metal layers (modelled by perfect electric conductors), with a total resonator diameter of 500 µm and a cavity height of 10 µm. The size of the holes is chosen to be on the order of the wavelength (20 µm). (b) Distribution of the holes for a device with 18% filling fraction. (c) Electric field distribution inside the cavity $|E_z|$ of a particular eigenmode, highlighted with a grey background in (d). (d) Near-field net-polarization $P_{Net}$ vs. far-field directionality (emission within an angle of $\Theta=10°$) for all eigenmodes obtained within a bandwidth of 400 GHz around 3.9 THz, indicating that, similar to the dipole-model, a near-field net-polarization is linked with the directional emission. (e) Calculated far-field of the particular mode highlighted in (d) with a grey background. (f) Incoherent superposition of 20 normalized far-fields corresponding to the modes with the highest near-field net-polarization $P_{Net}$ (red data points in (d)), resulting in a directional multi-frequency far-field.

Moreover, the electrically pumped nature of the used quantum cascade active region allows for highly controllable experimental conditions and is crucial for future applications of random lasers [30]. Our work thus opens up the way towards a new and versatile design principle for broadband coherent light sources.

**Funding.** Financial support received by the Vienna Science and Technology Fund (WWTF) through Project No. MA09-030, by the Austrian Science Fund (FWF) through Projects No. F25 (SFB IR-ON), No. F49 (SFB NextLite), No. W1210 (DK CoQuS), and No. W1243 (DK Solids4Fun). H.D. is an APART Fellow of the Austrian Academy of Sciences. The computational results presented have been achieved in part using the Vienna Scientific Cluster (VSC).

**Acknowledgment.** We thank G. Reider and M. Liertzer for very fruitful discussions.

# Supplementary Material

### Active region and fabrication

The active region is based on a three-well longitudinal-optical (LO) phonon depletion design [1] with a designed emission frequency around 3.9 THz. It was realized with GaAs/Al$_{0.15}$Ga$_{0.85}$As heterostructures, grown by molecular beam epitaxy. The layer sequence starting from the injector barrier in nanometers is **2.8**/8.5/**4.5**/5/<u>6.4</u>/5/**4.5**/8.5 (barriers in bold font). The center 6.4 nm of the 16.4 nm well is homogeneously n-doped with a density of $5.2 \times 10^{16}$ cm$^{-3}$ (underlined). The QCL module is repeated 222 times giving a total thickness of 10 µm. The active region was fabricated into a double-metal waveguide geometry with Au waveguide layers for vertical mode confinement, using a thermo-compression wafer bonding process. The resonators including the scattering elements were realized by structuring the top metal layer, acting as a self-aligned etch mask, using a photo-lithography/lift-off process and a subsequent reactive ion etching step, removing the active region material around the cavity and in the scattering elements. All devices were fabricated with a total cavity diameter of 500 µm in order to obtain a large emission aperture. The processed devices were mounted on a copper heat sink with indium and electrically contacted using a wire bonding technique.

### Measurement Setup

The lasers were mounted in a liquid helium cooled flow cryostat which is attached to a Bruker Vertex 80 Fourier transform spectrometer with a resolution of 2.25 GHz, collecting laser emission within an angle of 28°. The emitted radiation is measured using an integrated pyroelectric deuterated triglycine sulfate (DTGS) detector. Unless mentioned explicitly, all measurements were performed at a heat sink temperature of 5 K. For polarization sensitive measurements, a wire-grid polarizer is placed between the laser and the spectrometer. All measurements were performed in pulsed mode operation with a pulse length of 500 ns and a repetition rate of 80 kHz, modulated with 10 Hz in order to detect the signal with lock-in techniques. For the far-field measurements the cryostat is detached from the spectrometer and the emitted intensity is measured with a DTGS detector which is mounted on a 2D translation stage placed in 65 mm distance to the sample. Measurements of the absolute power were performed using a calibrated thermopile detector, mounted inside the cryostat, without any collimation optics. The obtained values were not corrected for the collection efficiency.

### Device Modelling

A simplified model based on the emission from discrete non-interacting oscillating dipole sources, is used to simulate the coherent emission from the random laser. Each dipole is located at the center of the corresponding hole $a_i$ and has a certain orientation of the (real-valued) dipole moment $\hat{p}$ in the $x,y$-plane and a constant temporal phase $\varphi_n$. For the calculations presented in Fig. 3, all dipoles are assumed to be in-phase $\varphi_n = 0$. In Fig. S4, the dipole-phases are chosen randomly in the range $-\frac{\pi}{2} < \varphi_n < \frac{\pi}{2}$. The electric field in the far-field limit ($kr \gg 1$) of a configuration with $N$ dipoles $\{p_n, a_n, \varphi_n\}$ is given by

$$E_\infty(r,t) = \sum_{n=1}^{N} \frac{k^2}{4\pi\varepsilon_0 r} (\hat{r} \times p_n) \times \hat{r} \, e^{ik(r-\hat{r}\cdot a_n)-i(\omega t+\varphi_n)}, \quad (S1)$$

where $r$ is an observation point in the far-field, $k$ is the wavevector and $\omega$ is the angular frequency. The radiant intensity (radiant flux per unit solid angle) is then given by

$$\frac{\partial \overline{P}}{\partial \Omega} = r^2 \langle S_r \rangle = \frac{r^2}{2Z} \|E_\infty\|^2 = \frac{k^4}{32\pi^2 \varepsilon_0^2 Z} \left\| \sum_{n=1}^{N} (\hat{r} \times p_n) \times \hat{r} \, e^{-i(k\hat{r}\cdot a_n + \varphi_n)} \right\|^2, \quad (S2)$$

where $S_r$ is the radial component of the Poynting vector and $Z$ is the impedance of free space.
The near-field net-polarization, important for directional emission, is quantified by

$$P_{Net} := \frac{1}{N}\left\|\sum_{n=1}^{N} \boldsymbol{p}_n e^{-i\varphi_n}\right\|, \qquad \|\boldsymbol{p}_n\| = 1, \qquad (S3)$$

where $N$ is the number of holes. If all dipole moments are oriented in the same direction and have the same temporal phase $\varphi_n$, $P_{Net} = 1$.

The results of the 3D finite-element model are obtained by solving the time-harmonic Maxwell equation for the electric field

$$(\nabla \times \nabla \times + n^2 k_m^2)\boldsymbol{E}_m(x) = 0, \quad (S4)$$

where $k_m$ are regarded as the eigenvalues of the differential equation and $\boldsymbol{E}_m$ as the eigenmodes of the system, commonly referred to as the quasi-bound states. In addition to that outgoing boundary conditions are imposed by employing a Cartesian perfectly matched layer (PML). The two metal waveguide layers (see Fig. 4(a)) were approximated by perfect electric conductors satisfying $\boldsymbol{E} \times \boldsymbol{n} = 0$ on the interface, where $\boldsymbol{n}$ is the normal vector to the surface. The active region was modeled as a dielectric with effective refractive index $n_{eff} = 3.6$. Note that the bonding wire was not incorporated in our model, due to its minor influence on the far-field [2].

The calculation of the electric field in the finite sized box confined by the PML was carried out using the open source library ngsolve, based on a high order finite element discretization technique, and the PARDISO LU-solver. The far-field was calculated using the Stratton-Chu formulas in the far-field limit with the open source boundary element library BEM++ [3]. We defined the near-field net-polarization, evaluated on the plane 1 µm above the top of the laser cavity, as

$$\boldsymbol{P}_{Net} = \frac{\sqrt{(\iint_{x,y} E_x(x,y)dx\,dy)\cdot(\iint_{x,y} E_x(x,y)dx\,dy)^* + (\iint_{x,y} E_y(x,y)dx\,dy)\cdot(\iint_{x,y} E_y(x,y)dx\,dy)^*}}{\iint_{x,y}\sqrt{E_x(x,y)\cdot E_x(x,y)^* + E_y(x,y)\cdot E_y(x,y)^* + E_z(x,y)\cdot E_z(x,y)^*}dx\,dy}, \quad (S5)$$

with the electric field components in $x, y$ and $z$ direction $E_{x,y,z}(x,y)$.

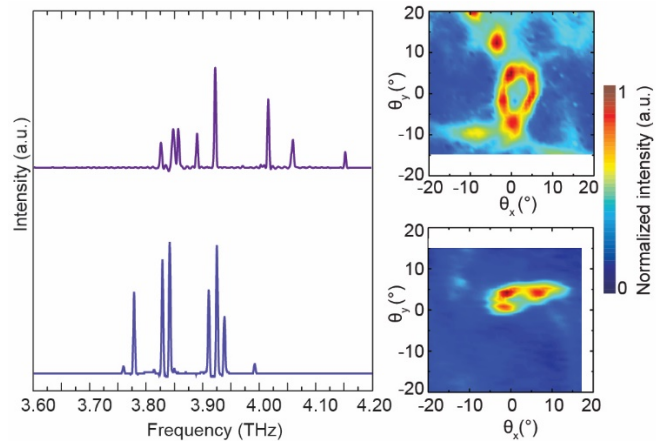

**Fig. S5**. Comparison of the measured far-fields and spectra of two random lasers with 34 % filling fraction. Due to the different hole arrangements, different modes and far-field patterns are obtained. However, as discussed in the main text, a general trend is observed that the number of modes and the far-field directionality is decreasing for increasing filling fraction.

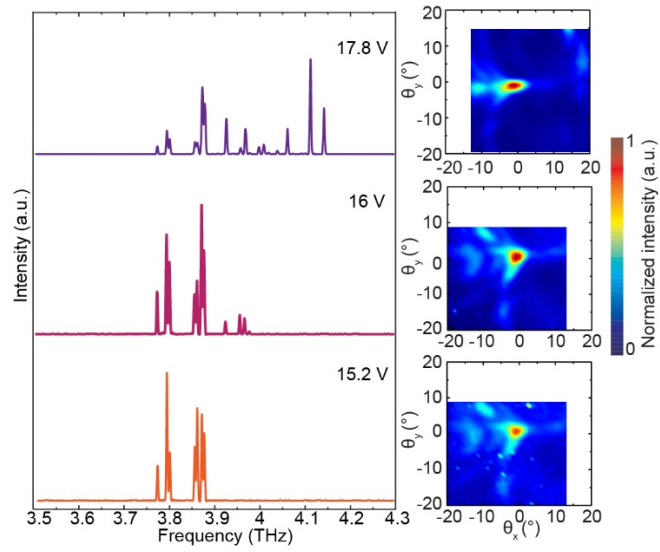

**Fig. S6.** Measured far-fields and spectral characteristics at different bias voltages. The measurements correspond to the device with 18 % filling fraction, presented in Fig. 2 in the main text. The far-fields and the spectra were recorded at heat sink temperature of 5 K at three different bias conditions (pump strengths). While the number of modes is increasing for increasing pump, the far-field quality improves in terms of less pronounced interference effects.

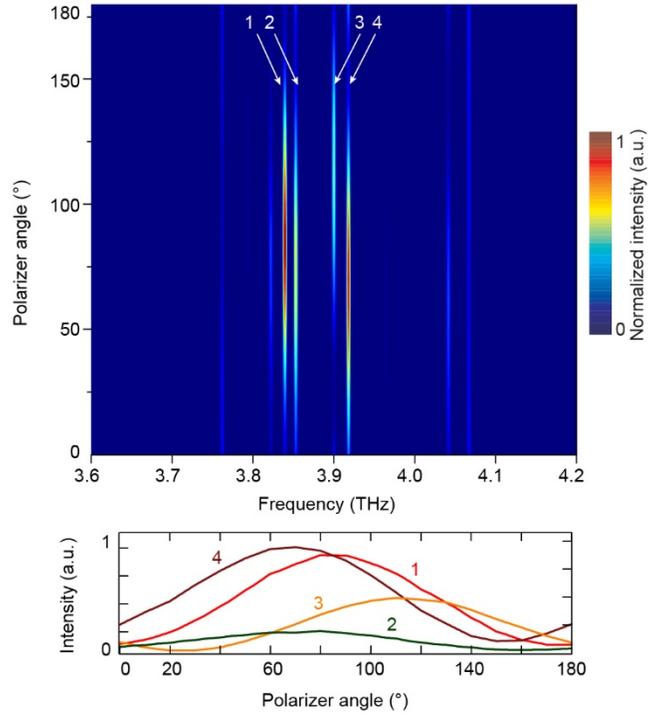

**Fig. S7.** Measured polarization of the individual spectral far-field components. For the measurement a wire grid polarizer is placed in between the cryostat window and the FTIR spectrometer to analyze the polarization properties of the far-field emitted perpendicular to the laser surface. Due to the optical arrangement, mainly the central lobe of the laser emission is measured. The individual laser modes are mainly linearly polarized since the signal almost vanishes for the polarizer arranged parallel to the mode polarization. Moreover, the polarization is different for the individual modes, implying an unequally oriented near-field net-polarization, which is in agreement with the dipole-model and the 3-dimensional calculations, discussed in the main text

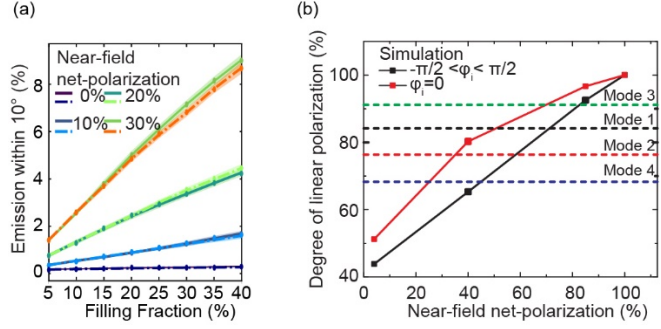

**Fig. S8.** Influence of the near-field phase on the far-field directivity. (a) The calculations were performed using the dipole model, discussed in the main text. The directivity is defined as the radiation emitted within an angle of 10° in surface direction. The results show a similar behavior of the directivity, regardless whether the dipoles are chosen to be in-phase ($\varphi_i = 0$, dashed lines) or if each dipole has a random (but constant) temporal phase $-\frac{\pi}{2} < \varphi_i < \frac{\pi}{2}$ (solid lines). (b) Degree of linear polarization vs. the near-field net-polarization for dipoles emitting in-phase and with random phase (35 % filling-fraction). The values corresponding to the measured modes in Fig. 3(f), and Fig. S3 are indicated by dashed horizontal lines, since the near-field net-polarization cannot be directly determined in the experiment. The degree of linear polarization is calculated using the Stokes parameters, $\Pi = \frac{\sqrt{S_1^2 + S_2^2}}{S_0}$, with $S_0 = P_{0°} + P_{90°}$, $S_1 = P_{0°} - P_{90°}$, $S_2 = P_{45°} - P_{135°}$ and the far-field intensity at different polarizer angles $P_{0°, 45°, 90°, 135°}$. The values were obtained for the far-field emission within 28°, corresponding to the geometry of the measurement setup. For in-phase dipole emission, a higher degree of linear polarization in the far-field is obtained.